\documentstyle[prl,preprint,aps,epsf]{revtex}
\begin{document}
\draft

\title{Magnetic--Field--Induced Exciton
Tunneling  in  Shallow Quantum Wells}

\author{A. Getter and I. E. Perakis}
\address{Department of Physics and Astronomy, Vanderbilt University, 
Nashville, Tennessee 37235}

\maketitle

\begin{abstract}

We study the  effect of the
magnetic field orientation 
on  the electroabsorption spectra of excitons 
confined in extremely shallow quantum wells.
When the applied electric field is parallel to the quantum 
well plane, we demonstrate that, for in--plane  magnetic field 
orientation,
the 
discrete confined exciton peak 
undergoes a transition 
into a 
continuum resonance.
In contrast, 
for perpendicular magnetic fields,
the exciton peak  exhibits  the usual 
Stark  red--shift. 
We show that such a dramatic   dependence 
on the magnetic field orientation originates from 
a resonant coupling 
between the confined and the bulk--like excitons.
Such coupling 
is caused by 
the interplay between the quantum--well potential
and a velocity--dependent two--body 
interaction between 
the exciton center--of--mass and relative motion
degrees of freedom
induced by the in--plane magnetic field. 
As a result, 
the exciton tunnels
out of the quantum well  as a whole 
without being ionized. 
We discuss possible experimental applications of this effect.

\end{abstract}


The response of semiconductor excitons and hydrogenic atoms 
to strong 
magnetic and 
electric fields 
has been the focus of much attention
in condensed matter, atomic, and molecular physics.
When the field--induced   forces  become comparable 
to the internal Coulomb interactions, 
the  response of such electron--hole (e--h) 
systems 
becomes  non--perturbative and provides 
valuable insight into many--body and confinement effects.
In addition to confining the e--h 
motion (Landau  quantization),  a 
magnetic field induces a
momentum--dependent  interaction between the 
center--of--mass
(CM) and relative (RM) e--h 
degrees of freedom
\cite{gorkov}.
Some experimental manifestations
of such two--body correlations 
were discussed in Ref. \cite{schme}
for atomic ions, 
and   in Ref. \cite{ilias} for 
shallow quantum well excitons.
In bulk semiconductors, the effect of an electric field is to 
ionize the RM Coulomb--bound state
and thus to 
broaden the
band--edge absorption spectrum 
(Franz--Keldysh effect) \cite{fke}.
In quantum wells (QW's), 
the confining 
 potential 
inhibits such an 
ionization for electric fields
perpendicular to the QW plane,  which results 
in an  exciton 
red--shift (Quantum Confined Stark Effect)
\cite{daniel}.
The latter effect   has become the 
basis of self--electrooptic--effect
switching devices
 \cite{miller}.
For efficient 
device operation, it is important to 
identify physical systems  
where 
an electric field
causes large changes in the  absorption 
spectrum 
(large contrast ratio).
In addition, in order 
to avoid a carrier  build--up 
within the QW region that would result into 
exciton broadening, 
it is important that
the optically--excited e--h pairs 
 rapidly escape from the QW. 
In  shallow QW's  with depth comparable 
to the bulk exciton binding energy,
it was shown that the combination 
of strong 
room--temperature excitons
and 
very short carrier escape 
times \cite{gero}
improves
the device switching 
speed
\cite{goos}.
Such considerations 
spurred recent efforts to study the special properties of 
shallow QW excitons  
in the crossover between 3D-- and 2D--like 
behavior.
This  confinement regime can be 
realized both in III--V 
\cite{bast,iotti,simm,brener,ilias}
and 
in II-VI \cite{2-6} 
semiconductor QW's.

In this Letter, we study the  
electroabsorption spectrum 
of magneto--excitons in extremely shallow QW's, 
whose 
depth is smaller than the 
Coulomb  binding energy.
Unlike in  typical electroabsorption experiments
\cite{ilias,daniel}, 
we consider weak electric fields $E$ {\em parallel to the QW 
plane}
and compare the exciton 
absorption lineshape 
for magnetic fields 
parallel ($B_{\|}$) 
or perpendicular 
($B_{\bot}$) 
to the QW plane.
We show that, with the same 
in--plane electric field, 
the discrete QW exciton peak 
undergoes a dramatic transformation if we turn on 
$B_{\|}$ 
perpendicular to the electric field
and gives way to a {\em continuum 
resonance}
with diminished absorption 
strength.
Such a strong dependence on the magnetic field orientation 
should be observable  in electroabsorption experiments 
for in--plane applied electric fields.
We attribute this transition 
 to a  strong 
two--body 
interaction between the CM and RM degrees of freedom
induced by $B_{\|}$. 
We show that, with $E \ne$0, 
this  leads to a 
resonant tunneling of the exciton as a whole 
out of the QW, 
without ionization of the  e--h 
Coulomb--bound state. 
We also discuss some possible applications of this effect. 

We start with the exciton Hamiltonian.
We choose the z--axis along the QW growth direction 
and the x--axis parallel to the in--plane 
electric field ${\bf E}$.
The  magnetic field is always  chosen perpendicular to the electric 
field, pointing either  along the 
z--axis
 ($ B_{\bot}$) 
or along the y--axis ($ B_{\|}$).
We work in the Landau gauge and 
denote by (${\bf R},{\bf P})$ 
and  $({\bf  r} = {\bf r_{e}} - {\bf r_{h}},{\bf p})$ 
the CM and RM position and momentum operators respectively.
A unitary transformation 
of the Hamiltonian 
defined by the operator $U= \exp(- i e {\bf r} \times {\bf B}/2 \hbar c)$ 
allows a partial  separation of the 
CM and RM degrees
of freedom \cite{gorkov}:
\begin{equation} 
H = H_{0}
+ V_{e}\left(Z + \frac{m_{h}}{M}z\right)
+ V_{h}\left(Z-\frac{m_{e}}{M}z\right),
\label{H}
\end{equation} 
where 
$M = m_{e} + m_{h}$
is the total exciton mass, 
$m_{e}$ and $m_{h}$ are the electron and hole masses,
$V_{e}$ and $V_{h}$ are  the electron and hole  
QW potentials respectively,
and the Hamiltonian $H_{0}$ is  
\begin{equation} 
H_{0}
=\frac{P_{Z}^{2}}{2 M} 
+ H_{\rm{RM}}
+ H_{\rm{int}}. \label{H0}
\end{equation} 
Here, the  
Hamiltonian 
$H_{\rm{RM}}$
describes  a RM 
quasiparticle in the presence of 
 one--body potentials 
due to the 
Coulomb interaction
and the electric and magnetic 
fields,
\begin{equation} 
H_{\rm{RM}}=
\frac{{\bf p}^{2}}{2 \mu} - \frac{e^{2}}{\epsilon r} 
+ \frac{e^{2}}{8 \mu c^{2}} 
({\bf B} \times {\bf r})^{2} 
+ \frac{e}{2 \mu c} \ \frac{m_{h} - m_{e}}{m_{e} + m_{h}} \ {\bf B}
\cdot {\bf L} + e \ {\bf E} \cdot  {\bf r},
 \label{RM}
\end{equation}
where  $\mu = m_{e} m_{h}/M$ is 
the reduced e--h mass, $\epsilon$ 
 the dielectric constant,  and ${\bf L}$ 
the angular momentum operator, while 
\begin{equation} 
H_{\rm{int}} = \frac{e}{cM} 
\left( {\bf P} \times {\bf B} \right) \cdot {\bf r} 
\label{Hint}
\end{equation} 
describes a 
{\em two--body} 
interaction 
between the CM and RM degrees of freedom.
Since we are concerned 
with extremely shallow QW's, 
we assume  the bulk exciton  bandstructure 
and  consider
a single  valence band with $m_{h}=0.15  \ m_{0}$, 
 $m_{0}$ being the free electron mass, and 
a conduction band with mass
$m_{e} = 0.067 \ m_{0}$.

 In the absence of the QW potential, 
the CM and RM subsystems are separable and 
the two--body exciton problem 
can be 
reduced to two one--body problems.
Since  in this case 
the CM and RM motions are 
uncorrelated,
the exciton wave function
is a product of 
a  CM and a RM quasiparticle contribution. 
One might therefore expect 
weak CM--RM correlations 
in  extremely shallow 
QW's (with 
depth smaller 
than the Coulomb binding energy),
in which case 
a factorized 
(adiabatic) 
 exciton wave function 
\cite{ilias,2-6}
would  provide  a good approximation. 
However, as discussed in Ref. \cite{ilias}, 
such an adiabatic approximation fails 
in the absence of translational invariance
due to 
a strong
 CM--RM correlation induced 
by $H_{\rm{int}}$.
This is due to 
the existence
(for $H_{\rm{int}} \ne 0$)
 of 
small--energy excitations 
between the ground states of 
$H_{\rm{RM}} + H_{\rm{int}}$
corresponding 
to  different CM--momentum values.
In particular, 
in the absence of translational invariance,
$P_{Z}$ is no longer a constant of motion, 
so that the CM motion can 
excite 
the above low--energy RM degrees of freedom, 
which in turn affect strongly the CM motion. 
For $E$=0, 
such CM--RM correlation effects on the discrete 
exciton 
ground state
can be described by using 
the  general variational 
wave function
of Ref. \cite{ilias}.
However, for $E,  B_{\|} \ne 0$, 
we find that 
such a  calculation 
does not converge, indicating 
that  the QW--confined  exciton state is 
no  longer the 
ground state of the system. 

We proceed by 
expanding the two--body exciton 
wave function $\Phi(Z, {\bf r})$
in the basis of eigenstates of the bulk Hamiltonian
$H_{0}$ for  different values of $P_{Z}$: 
\begin{equation} 
\Phi(Z, {\bf r})=
\sum_{P_Z} \ \Psi(P_Z) \ e^{iP_Z Z/\hbar} \
\phi_{P_Z}({\bf r}),
\label{wav}
\end{equation} 
where 
$\phi_{P_Z}({\bf r})$ 
is the ground state of $H_{0}$ 
for a given value of $P_{Z}$.
The  exciton wave function 
Eq.\ (\ref{wav})
takes into account  excitations between 
the low--lying 
RM
 ground states  of $H_{0}$ 
corresponding to different
values of 
 $P_{Z}$. These are induced 
by the CM motion and are neglected in the adiabatic approximation.
It is assumed
that, 
due to the
fact that 
the 
finite 
exciton 
binding energy
is larger than the QW depth, 
the 
mixing
of the 
excited magneto--exciton states 
for a given $P_{Z}$ 
is weak.
For $E$=0 or $B_{\|}=0$, 
the above wave function compares very well 
with the
variational solution 
of Ref. \cite{ilias}
in extremely shallow QW's.

Using the wave function 
Eq.\ (\ref{wav}), 
the Schr\"{o}dinger equation with the total Hamiltonian 
$H$ leads to the following equation 
for  the CM--momentum
wave function
$ \Psi(P_{Z})$:
\begin{equation} 
\left[ \varepsilon(P_{Z})- \varepsilon \right] \Psi(P_{Z}) 
= - \sum_{P_{Z}^{'}} \ V_{\rm{eff}}(P_{Z},P_{Z}^{'}) \
\Psi(P_{Z}^{'}) 
\label{CM}
\end{equation} 
where $\varepsilon$ is the exciton energy, 
$ \varepsilon(P_{Z})$ is the
ground state energy
of the 
Hamiltonian $H_{0}$ for a given value of  $P_{Z}$,   and 
\begin{equation} 
V_{\rm{eff}}(P_{Z},P_{Z}^{'}) 
= \int dZ \ e^{i (P_{Z}^{'}-P_{Z}) Z /\hbar} 
\ \langle \phi_{P_{Z}} | V_{e}\left(Z + \frac{m_{h}}{M}z\right)
+ V_{h}\left(Z-\frac{m_{e}}{M}z\right) 
| \phi_{P_{Z}^{'}} \rangle \label{V} 
\end{equation} 
is an effective non--local QW potential 
that depends on both the 
RM wave functions
and the CM momentum. 
To calculate $ \varepsilon(P_{Z})$ and 
$V_{\rm{eff}}$, we 
diagonalized
the RM Hamiltonian $
H_{0}$ 
for different values of $P_{Z}$ using 
a real--space Gaussian 
basis set similar to Ref. \cite{ilias} and 
including basis 
elements with all 
the  prefactors  allowed by the reduced 
symmetry when $E \ne 0$. 
Although for $B_{\|}=0$ 
Eq.\ (\ref{CM}) has a discrete  ground state, 
with finite 
$B_{\|}$ and $E$
  we found   many  eigenstates 
closely--spaced in energy, suggesting that the 
breakdown of adiabaticity changes the 
qualitative behavior of the 
CM motion.

With the above--obtained  exciton wave function 
$\Phi(Z, {\bf r})$, 
we  calculated the absorption spectrum 
using 
Fermi's golden rule.
Our results  
in the frequency range of the 
confined exciton 
are presented in
Fig.\ \ref{fig1}(a),
for magnetic field 
perpendicular to the QW plane 
($B_{\bot}$),
and in  Fig.\ \ref{fig1}(b), 
for magnetic field parallel 
to the QW plane ($B_{\|}$).  
Both magnetic field orientations 
are perpendicular to that of the  
in--plane electric field. 
Despite  the fact that 
the QW depths 
$V_{e}$=1.2meV and $V_{h}=$0.8meV
are much smaller  than the bulk exciton binding energy, 
$E_{B} \sim$ 4meV,
our calculations show 
a
sharp contrast
for the two different magnetic field 
orientations
between  the changes of the  
magneto--exciton lineshape  
induced by the electric 
field. 
With  $B_\|=0$, 
we obtain 
a small 
Stark  red--shift of the exciton peak [see Fig.\ \ref{fig1}(a)].
With finite $B_{\|}$ however, 
the same  weak in--plane 
electric 
field
induces a large exciton broadening 
and 
decrease 
in absorption strength [see Fig.\ \ref{fig1}(b)].
It is important to note here that 
such an effect 
is not due to 
the ionization of the RM exciton  state. 
In fact, since both 
$B_\|$ and $B_\bot$ 
are perpendicular 
to the electric field
pointing along the x--axis, 
the {\em same} 
diamagnetic potential, 
{\em quadratic} in $x$, 
inhibits the 
RM ionization 
by opposing the 
electric field 
potential, 
which is {\em linear} in $x$. 
Unlike in the $ B$=0 case, the 
electric field 
reduces the  Coulomb 
binding energy without 
ionizing  
the  magneto--exciton.

To interpret the above dramatic 
difference in the electroabsorption spectrum, 
 let us compare the 
exciton  Hamiltonians for   the two different magnetic field 
orientations. 
The differences in $H_{\rm{RM}}$ 
cannot explain our effect, 
which is caused by 
the two--body interaction 
$H_{\rm{int}}$. 
In the bulk semiconductor, 
the CM momentum 
${\bf P}$ is a constant of motion. For 
optically--active excitons, 
$ {\bf P}=0$ in the dipole approximation
and therefore 
$H_{\rm{int}}=0$ 
does not affect the linear optical spectra. 
In a   QW however, 
$P_{Z}$ becomes a dynamical variable due to the breakdown 
of the translational invariance. 
For $ B_\| =0$, 
$H_{\rm{int}}$ only depends 
 on the in--plane
components of the CM
momentum, which 
vanish  for  optically--active
excitons. 
Therefore, in this case, 
  $H_{\rm{int}}$ does not affect 
the absorption spectrum.
On the other hand, 
with finite $B_\|$, one has 
$H_{\rm{int}}=-  eB_{\|} x  P_{Z}/cM$.
This two--body Hamiltonian  may be thought of as 
a {\em fluctuating}  electric field 
potential, which
intimately couples the CM and RM degrees of freedom. 
By changing the magnetic field orientation, 
we     therefore tune the  strength
of $H_{\rm{int}}$, which 
affects the exciton 
CM motion via the  renormalization 
of $\varepsilon(P_{Z})$ and $V_{\rm{eff}}$
[see Eq.\ (\ref{CM})]. 

Let us first consider the dispersion relation
$\varepsilon(P_{Z})$ of the CM degree of freedom. 
With 
magnetic field perpendicular to the QW plane,
the  CM--RM interaction  $H_{\rm{int}}$
vanishes and the 
momentum--dependence 
of $\varepsilon(P_{Z})$ is quadratic.
This  is no longer the  case  when 
both $ B_\|$ and $ E$ are finite.
In  Fig.\ \ref{fig2}(a) 
we show the dispersion relation 
for in--plane magnetic field $ B_\|$ 
and  different values of the  electric field.
Our numerical calculation 
is consistent with the 
analytic asymptotic expressions 
derived in Ref. \cite{gorkov} 
for  a bulk semiconductor 
and  high magnetic fields.
The total 
effective electric  field 
${\cal E}(P_{Z}) = 
E -  B_\| P_{Z} /c M$
acting on the RM depends on the CM momentum, which in QW's 
is no longer a constant of motion.
For  values of $P_{Z}$ corresponding to 
large ${\cal E}(P_{Z})$,
such an 
electric field 
dominates over 
the Coulomb interaction and 
leads to a linear dependence of 
 $\varepsilon(P_{Z})$
on the CM momentum. 
On the other hand, 
for values of $P_{Z}$ 
corresponding to small 
${\cal E}(P_{Z})$,
the Coulomb interaction 
leads to a 
{\em local  
minimum}
in the 
CM dispersion relation. 
The most important feature of 
the  spectrum of  Fig.\ \ref{fig2}(a)
is the
Coulomb--induced 
 {\em degeneracy} between 
high momentum 
eigenstates and 
those corresponding to $P_{Z}$ 
close to the dispersion minimum.

Let us now turn to  the effective QW
 potential $V_{\rm{eff}}$.
As can be seen from Eq.\  (\ref{CM}), 
$V_{\rm{eff}}$ 
causes an appreciable mixing 
of all 
the 
low--energy
ground
 states 
of the RM Hamiltonian 
 $H_{\rm{RM}} + H_{\rm{int}}$ 
corresponding to the 
different 
values of   $P_{Z}$.
Such an excitation of the RM  
is absent within the 
adiabatic approximation and 
changes  the qualitative behavior of the 
CM momentum wave function 
$\Psi(P_{Z})$. Indeed, 
for finite  $B_{\|}$ and $ E$, 
we find 
many  exciton eigenstates 
with energies closely--spaced 
around the energy  of 
 the  confined 
exciton.
The 
probability densities for some 
of these optically--active states 
are shown in Fig.\ \ref{fig2}(b).
As can be seen,
the CM state  is a superposition of
two states,
 one 
with  wave function 
with finite momentum distribution 
centered 
at small  positive 
$P_{Z}$, 
and 
another 
with  wave function  
sharply peaked at large $P_{Z}$.
The first peak in 
$\Psi(P_{Z})$ corresponds to the QW--confined exciton state, 
whose energy lies below the local minimum of 
$\varepsilon(P_{Z})$, 
while the second sharp peak comes from 
the coupling of bulk excitons whose energies
are degenerate with that of the discrete confined state.
The above states are resonantly coupled 
by $V_{\rm{eff}}$, meaning that the 
confined exciton can tunnel into the continuum 
of bulk excitons. 
Such an effect
for $B_{\|} \ne 0$  
manifests itself 
via the transformation of the discrete QW exciton into 
a continuum resonance [see Fig.\ \ref{fig1}(b)].
On the  other hand, for $ B_\|=0$, 
there is no degeneracy 
and  we obtain 
a  sharp  QW--confined exciton peak [see Fig.\ \ref{fig1}(a)]. 

Let us  
now 
discuss some possible applications 
of our results. 
First, a magnetic field of a few Tesla allows 
one to strongly modify 
the exciton absorption by 
using 
very weak electric fields. 
For example, as shown in 
Fig.\ \ref{fig1}(b), 
an  electric field  of only $ E \sim$ 3~kV/cm
leads to a 
$\sim$80\% decrease 
in  the exciton  strength.
Such a sharp contrast ratio is desirable 
for the efficient operation 
of switching devices
\cite{goos,miller}.
Unlike in typical 
electroabsorption experiments 
 \cite{goos,miller,daniel},
 sharp changes in the absorption spectrum 
are  achieved by using a weak in--plane electric field
that does not ionize the RM 
exciton state. This 
suppresses 
 the undesirable 
Franz--Keldysh broadening
that limits the contrast ratio. 
In our case, the confined exciton {\em as a whole} 
tunnels out of the QW 
and into degenerate {\em dark} 
bulk--like exciton states with high CM momentum.
This 
as well as 
 the fact that the 
electric field shifts the minimum of $\varepsilon(P_{Z})$ 
to nonzero CM momentum values 
lead to a 
very strong decrease in the exciton absorption strength.

In conclusion, 
we showed that, by changing the magnetic field 
orientation from perpendicular to parallel 
to the plane of an extremely shallow quantum well,
the discrete confined exciton state transforms into a continuum 
resonance 
in the presence of 
a weak in--plane 
electric field. 
This transition is due to  a resonant 
tunneling of the confined exciton as a whole 
out of the QW region, 
without ionization of the RM bound state. 
This is 
caused 
by  the interplay between  the quantum well potential
and a two--body velocity--dependent interaction between the 
CM and RM degrees of freedom
induced with an in--plane magnetic field.
The above dramatic effect can be  observed  
with electroabsorption 
experiments 
that apply in--plane electric fields to 
extremely shallow 
GaAs/AlGaAs QW excitons. 
The role 
of shallow QW potentials
and the  
above 
interactions on 
the transition from regularity to chaos 
\cite{schme} 
and the dynamical exciton--exciton interactions 
in ultrafast four--wave--mixing spectroscopy 
\cite{chemla} 
remain to be studied.

This work was supported by the NSF CAREER award ECS-9703453, and, in
part, by ONR Grant N00014-96-1-1042  and by Hitachi Ltd.
Part of this work was performed 
while A.G. was at Rutgers University. 
We thank 
W. Knox, K. W. Goossen, J. E. Cunningham, T. V. Shahbazyan, 
and especially S. A. Jackson for valuable discussions.

\begin{figure}
\caption{Change in the 
QW--confined magneto--exciton 
absorption 
lineshape  induced by  an in--plane electric field
of 
$E$=2.9 kV/cm for
(a) 
perpendicular magnetic field 
$B_{\bot}$=10T,
and (b) in--plane  magnetic field 
$B_{\|}$=10T.
} \label{fig1}
\end{figure}

\begin{figure}
 \caption{ (a) CM  dispersion relation
$\varepsilon(P_{Z})$ 
for  $B_\| = 10~T$ 
and E=0 (solid curve),
1.16KV/cm (dashed  curve), 
1.74 KV/cm (long--dashed curve), 
2.32 KV/cm (dotted--dashed curve), 
and 2.9 KV/cm (dotted curve). 
(b) Momentum probability density $| \Psi(P_{Z}) |^{2}$
of several low--lying  QW exciton 
states
for   $B_\| = 10$~T and $E$=2.9KV/cm} 
\label{fig2}
\end{figure}

\clearpage 

\epsfxsize=6.0in
\epsffile{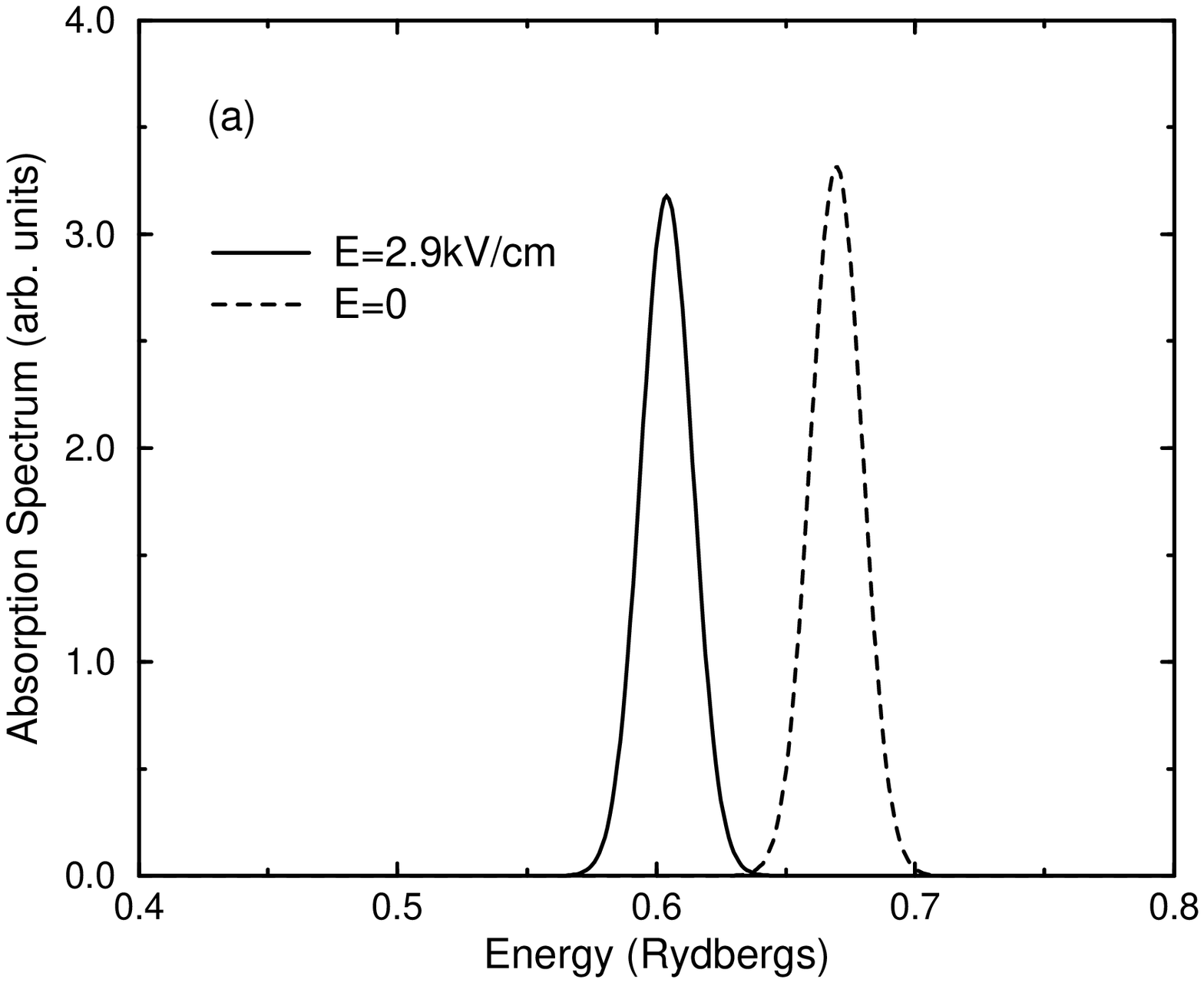}
\vspace{3in}
\centerline{Fig. 1}

\epsfxsize=6.0in
\epsffile{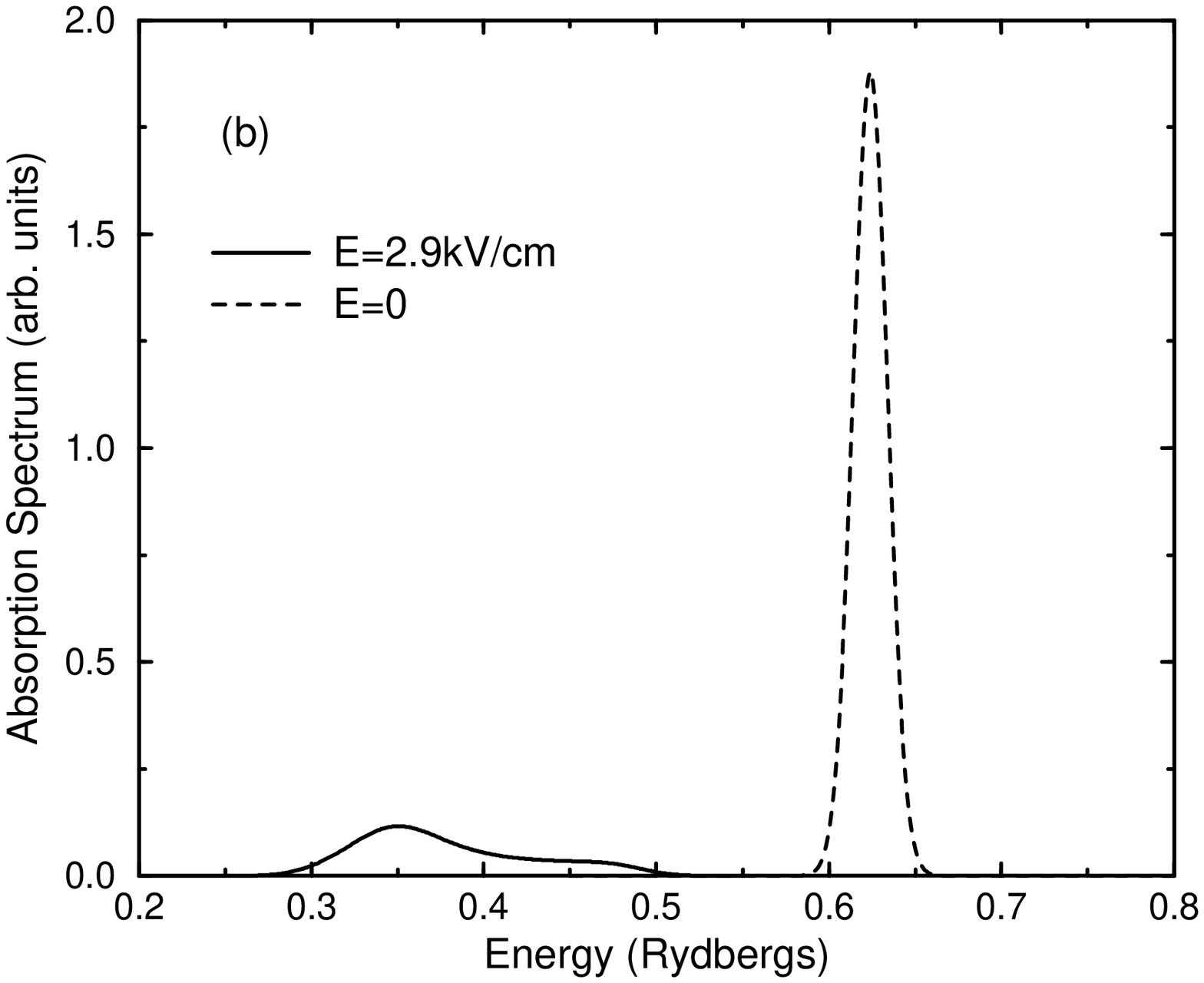}
\vspace{3in}
\centerline{Fig. 1}

\epsfxsize=6.0in
\epsffile{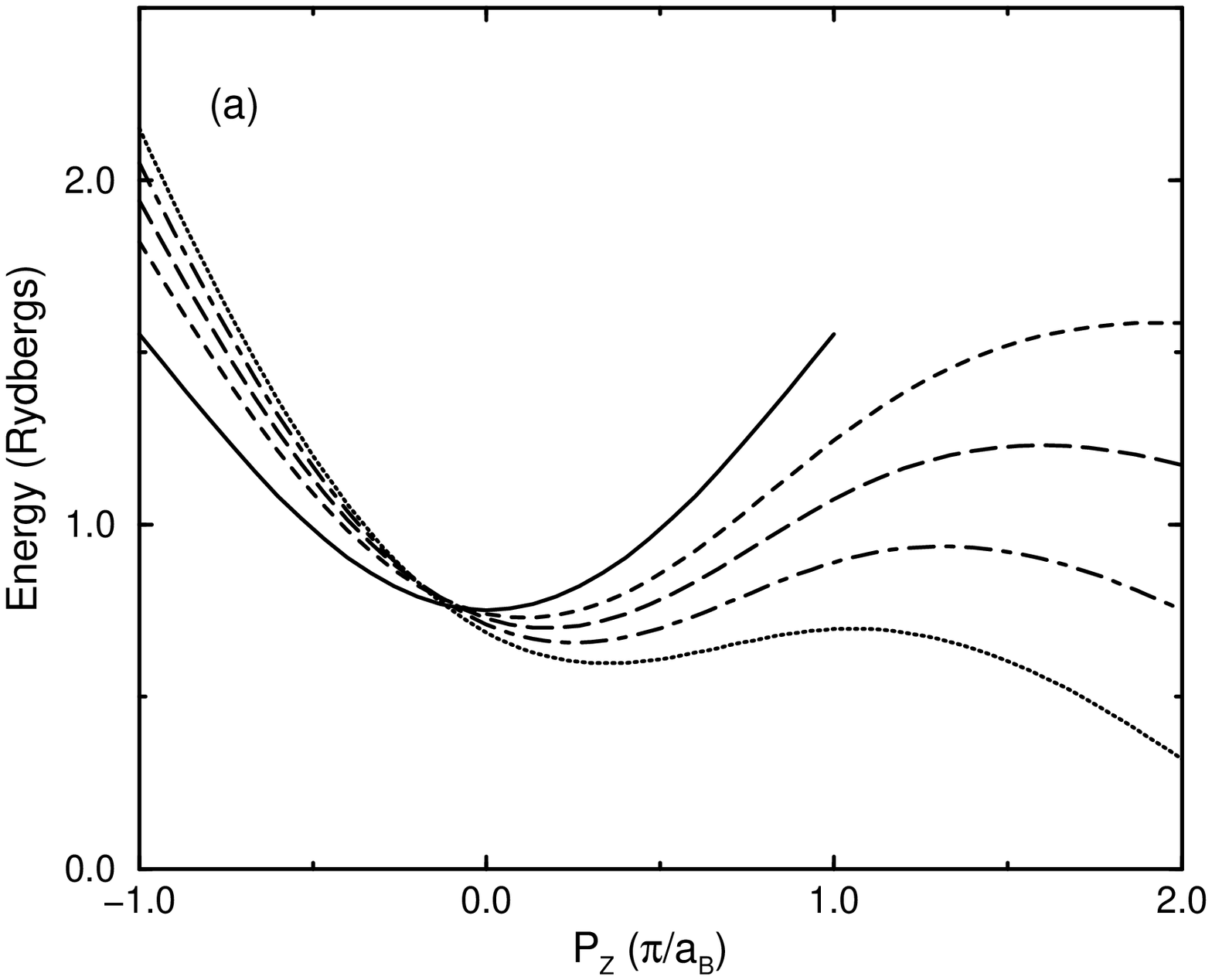}
\vspace{3in}
\centerline{Fig. 2}

\epsfxsize=6.0in
\epsffile{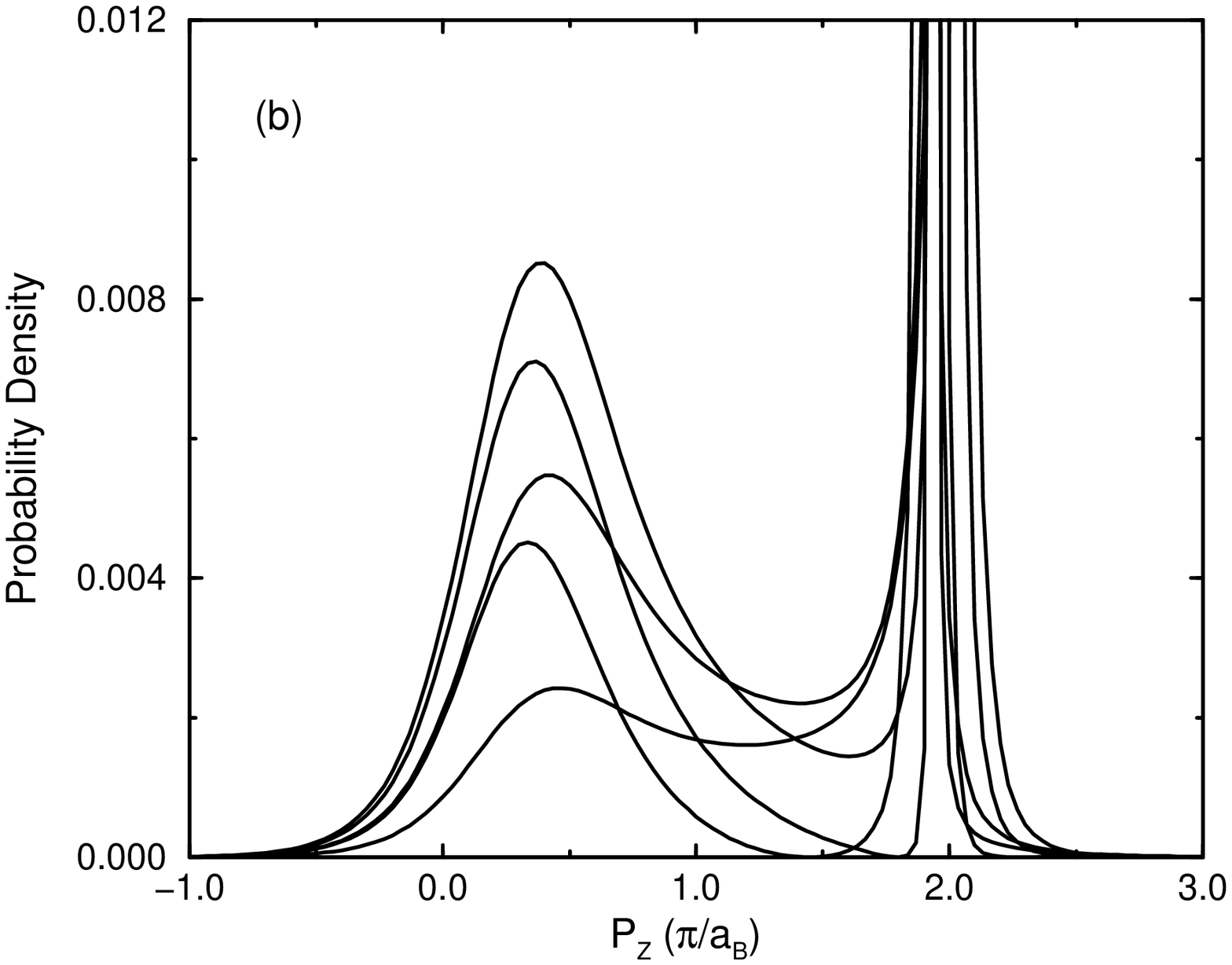}
\vspace{3in}
\centerline{Fig. 2}

\end{document}